\documentclass[apj]{emulateapj}
\usepackage{float}
\usepackage{epsfig}
\usepackage{graphicx}
\usepackage{graphics}
\usepackage[latin1]{inputenc}
\usepackage{latexsym}
\usepackage{footnote}
\usepackage[bottom]{footmisc}
\newcommand{\cl}{\textit{clear}}
\newcommand{\nd}{{\nodata}}

\begin{document}
\submitted{Accepted for Publication in the Astronomical Journal}
\title{Difference Image Analysis of Defocused Observations with CSTAR}

\author{Ryan J.~Oelkers\altaffilmark{1,2,*}, Lucas M.~Macri\altaffilmark{1}, Lifan Wang\altaffilmark{1,3,4}, Michael C.~B.~Ashley\altaffilmark{5}, Xiangqun Cui\altaffilmark{4,6}, \\
Long-Long Feng\altaffilmark{3,4}, Xuefei Gong\altaffilmark{4,6}, Jon S.~Lawrence\altaffilmark{5,7}, Liu Qiang\altaffilmark{4,8}, Daniel Luong-Van\altaffilmark{5}, Carl R.~Pennypacker\altaffilmark{9}, \\
Huigen Yang\altaffilmark{4,10}, Xiangyan Yuan\altaffilmark{4,6}, Donald G.~York\altaffilmark{11}, Xu Zhou\altaffilmark{4,8}, Zhenxi Zhu\altaffilmark{3,4}}

\altaffiltext{1}{George P.~and Cynthia W.~Mitchell Institute for Fundamental Physics and Astronomy, Department of Physics and Astronomy Texas A\&M University, College Station, TX 77843, USA}
\altaffiltext{2}{2012 East Asian and Pacific Summer Institutes Fellow}
\altaffiltext{3}{Purple Mountain Observatory, Chinese Academy of Sciences, Nanjing, China}
\altaffiltext{4}{Chinese Center for Antarctic Astronomy, Nanjing, China}
\altaffiltext{5}{School of Physics, Univ. of New S. Wales, NSW, Australia}
\altaffiltext{6}{Nanjing Institute of Astronomical Optics and Technology, Nanjing, China}
\altaffiltext{7}{Australian Astronomical Observatory, NSW, Australia}
\altaffiltext{8}{National Astronomical Observatories, Chinese Academy of Sciences, Beijing, China}
\altaffiltext{9}{Institute of Nuclear and Particle Astrophysics, Lawrence Berkeley National Laboratory, Berkely, CA, USA}
\altaffiltext{10}{Polar Research Inst. of China, Pudong, Shanghai, China}
\altaffiltext{11}{Department of Astronomy and Astrophysics and Enrico Fermi Institute, University of Chicago, Chicago, IL, USA}
\altaffiltext{*}{Corresponding author, {\tt ryan.oelkers@physics.tamu.edu}}

\begin{abstract}
The Chinese Small Telescope ARray (CSTAR) carried out high-cadence time-series observations of 27 square degrees centered on the South Celestial Pole during the Antarctic winter seasons of 2008, 2009 and 2010. Aperture photometry of the 2008 and 2010 \textit{i}-band images resulted in the discovery of over 200 variable stars. Yearly servicing left the array defocused for the 2009 winter season, during which the system also suffered from intermittent frosting and power failures. Despite these technical issues, nearly 800,000 useful images were obtained using \textit {g, r} \& {\cl} filters. We developed a combination of difference imaging and aperture photometry to compensate for the highly crowded, blended and defocused frames. We present details of this approach, which may be useful for the analysis of time-series data from other small-aperture telescopes regardless of their image quality. Using this approach, we were able to recover 68 previously-known variables and detected variability in 37 additional objects. We also have determined the observing statistics for Dome A during the 2009 winter season; we find the extinction due to clouds to be less than 0.1 and 0.4 mag for $40\%$ and $63\%$ of the dark time, respectively. 
\end{abstract}

\section{Introduction}

Time-series photometry has long been one of the main tools to study many problems in astrophysics. Over the last decade, technological advances have enabled a large increase in the number of nearly-uninterrupted, high-quality and high-cadence observations which have resulted in an increased understanding of stellar astrophysics, the discovery of hundreds of exoplanets and the detection of rare transient events \citep{Baglin2006, Borucki2010, Law2010}. Many scientific teams have deployed arrays of small aperture telescopes to study time-series phenomena because they are are relatively inexpensive and highly reproducible \citep{Bakos2002, Pollacco2006, Pepper2007}. 

Unfortunately, small telescopes can suffer from a large number of systematics not found in their larger counterparts. Small telescopes typically have large fields of view ($20-100$~sq. deg) which lead to large pixel scales ($>6-15\arcsec/$pix). This guarantees many sources will be blended and most environments will be crowded. Smaller optics lead to higher vignetting and positional variations in the point spread function (PSF) across the detector, requiring more complex photometric reduction procedures. Despite these disadvantages, many small aperture telescopes have produced high quality photometry. 

The Chinese Small Telescope ARray (CSTAR) was designed to test the feasibility and quality of an observatory stationed at Dome A on the Antarctic Plateau. Dome A is considered to be one of the most promising observing sites on Earth with low temperature, high altitude (4200 m), extremely stable atmospheric conditions ($<0.4$~mag extinction for 70\% of the time) and nearly-uninterrupted dark conditions for 6 months \citep{Zou2010, Zhou2010, Wang2011, Wang2013}. 

CSTAR was deployed at Dome A during the 2008, 2009 and 2010 Antarctic winter seasons. Previous studies of the photometry using aperture photometry from the 2008 and 2010 winter seasons have shown remarkable clarity, coverage and precision \citep{Zou2010, Zhou2010, Wang2011, Wang2013}. More than 200 variable stars with $i<15.3$~mag were categorized including exoplanet candidates, Blazhko-effect RR Lyraes, a possible type II Cepheid in an eclipsing binary system and a star with regular milli-magnitude variations on extremely short timescales \citep{Wang2011, Wang2013, WangS2014}.

This paper presents the first complete analysis of the data from the 2009 Antarctic winter season, using a combination of previously known techniques to deal with the defocused images obtained during that period. Our methodology can be applied to observations from other small telescopes regardless of the PSF shape and will specially benefit imaging of crowded fields. The rest of our paper is organized as follows: \S2 describes the 2009 CSTAR observations; \S3 details the data processing steps; \S4 presents our photometric reduction process; \S5 analyzes the photometric noise of our reduction procedure; \S6 discusses the variable stars in our field; \S7 contains our conclusions.

\medskip

\medskip

\section{Observations}
  
CSTAR was deployed at Dome A in early 2008 and carried out observations during three Antarctic winter seasons before returning to China for comprehensive upgrades in early 2011; the following description applies to the original version of the system. It is composed of four Schmidt-Cassegrain wide-field telescopes, each with a 145mm aperture and a field of view $4.5^\circ$ on a side. The focal planes contain ANDOR DV435 1K$\times$1K frame-transfer CCDs with a pixel size of 13$\mu$m, equivalent to a plate scale of $15\arcsec/$pix. Filters are mounted at the top of the optical tubes, with a 10W electric current run through a coating of indium tin oxide to prevent frosting \citep{Yuan2008}. Three of the filters are standard SDSS \textit{gri} while the remaining one is a clear filter (hereafter, {\cl}). CSTAR contains no moving parts and the telescopes do not track. In order to keep the resulting drift from subtending more than a pixel, the telescopes are pointed towards the South Celestial Pole (hereafter, SCP) and exposures are kept below 30s. The observations presented in this paper began on 2009 March 20, with the exposure time set to 5 seconds. This was increased to 20 seconds on 2009 April 14 as the sky level decreased.

Routine servicing of the system in early 2009 inadvertently left all telescopes out of focus to varying degrees. CSTAR\#1 (fitted with the $i$ filter) failed to return any data. CSTAR\#2 (\textit{g}) had a somewhat regular, torus-like PSF. CSTAR\#3 ({\cl}) had the best overall focus but was plagued by intermittent frosting of the lens. CSTAR\#4 (\textit{r}) had an irregular, torus-shaped PSF. Figure~\ref{fig:ref} shows $200\times100$~pix subsections of images obtained with each telescope to show the extent of the defocusing. Table~\ref{tb:log} lists the number of useful images acquired and the date of final power loss for each telescope.

\section{Data Pre-Processing}

\subsection{Flat Fielding and Bias Subtraction}
The first step in our pre-processing was the subtraction of a bias frame and the generation of an accurate flat field. We used bias frames obtained during instrument testing in China \citep{Zhou2010a} while sky flats were generated from our observations. We selected $\gtrsim 3000$ frames where the sky background was higher than 7,000 and 4,000 ADU for the \textit{g} and \textit{r} flats, respectively; the corresponding values for {\cl} were $\gtrsim 8000$ frames and 10,000 ADU. We bias-subtracted, scaled and median-combined the selected frames to make a temporary flat field, applied it to the images, masked any detected stars and repeated the process to generate the final flat fields.

\begin{figure}[t!]
\centering
\includegraphics[width=85mm]{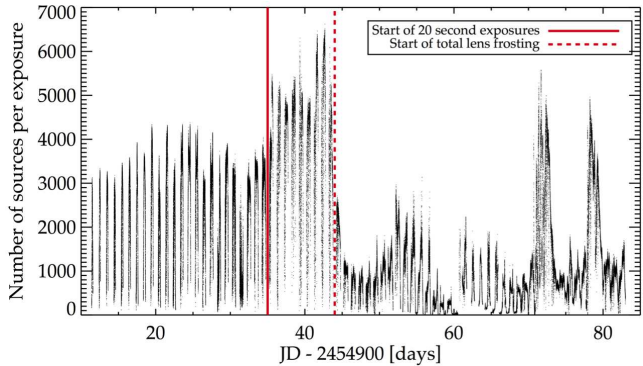}
\caption{The number of detected sources with the DAOPHOT, FIND (Stetson, 1987) routine for {\cl} as a function of Julian date. The number of stars per exposure remains somewhat constant until $\sim$ 10 days after the switch from 5s exposures to 20s exposures signifying filter frosting.\label{fig:frost}}
\end{figure}

During this process, we found transient structures in the {\cl} images, which we determined to be the result of partial-to-complete frosting of the filter. Quantitatively, this can be seen in the variation of the number of sources recovered in each frame after flat fielding, as shown in Figure~\ref{fig:frost}. The number of sources increases once the exposure time is increased from 5 to 20 seconds, 25 days after the start of observations. The number of stars drops dramatically $\sim10$~days later, signaling the advanced stages of filter frosting. We removed $\sim 40$\% of all {\cl} frames after JD 2454945.0, when the star counts dropped below 3000/frame. We later removed another $\sim 25$\% of the remaining {\cl} frames in which all light curves exhibited a significantly higher dispersion ($\sim 0.2$~mag or greater), which we interpreted as evidence of intermittent frosting (see Figure~\ref{fig:frost} for details).

\begin{figure}[t!]
\centering
\includegraphics[width=70mm]{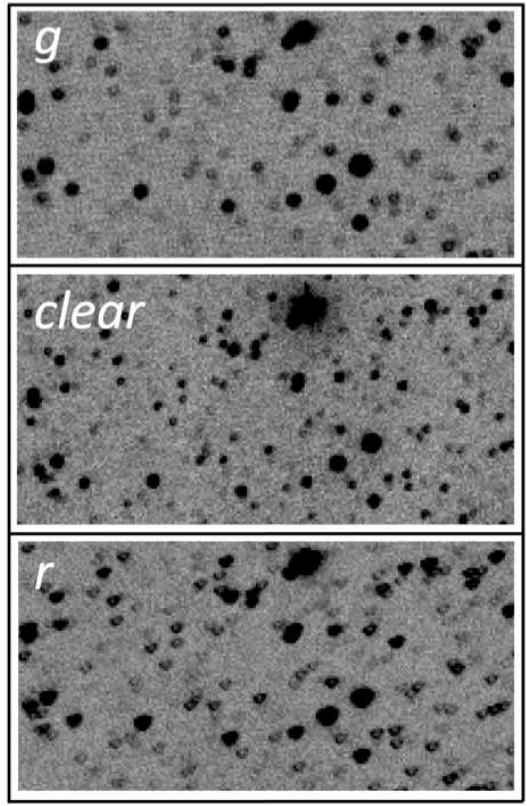}
\caption{$200 \times 100$~pixel subsections of each reference frame in \textit{g}(top), {\cl} (middle) and \textit{r} (bottom), all centered on the south celestial pole (SCP). Each frame shows the varying level of defocusing in each telescope, giving rise to the donut-shaped PSFs. Color has been inverted for clarity.\label{fig:ref}}
\end{figure}

\begin{deluxetable}{lcr}
\tablewidth{0pt}
\tablecaption{Observation Log\label{tb:log}}  
\tablehead{\colhead{Band} & \colhead{Number of} & \colhead{Date of total} \\ 
\colhead{}      & \colhead{useful images} & \colhead{power loss}}
\startdata
\textit{g} & 241,903 & 2009 June 1 \\ 
\cl &  74,010 & 2009 May 31 \\ 
\textit{r} & 483,109 & 2009 July 30 
\enddata
\end{deluxetable}

\subsection{Background \& Electronic Pattern Subtraction}
After flat fielding, some frames still exhibited a low-frequency residual background, likely due to moonlight or aurora. We applied a residual background subtraction following the approach of \citet{Wang2013}. The residual background model is constructed by sampling the sky background every $32\times 32$ pixels over the entire detector. Bad or saturated pixels are excluded from each sky sample. A model sky is then fit inside each box and interpolated between all boxes to make a thin plate spline \citep{Duchon1976}. We used the IDL implementation GRID$\_$TPS to make the spline which is subtracted from the frame. 

Images in all bands exhibited a similar electronic noise pattern which became significant at low sky background levels. We used the following procedure to remove the pattern. All stars at $2.5\sigma$ above the sky level and all bad pixels were masked and replaced with random values based on a Gaussian distribution that matched the properties of the background. 

We calculated the Fast Fourier Transform (FFT) of each image and identified significant peaks with a power greater than $10^{-3}$, which corresponded to the frequency of the electronic pattern. We generated an image containing the unwanted pattern by taking the inverse FFT of the selected frequencies, which was subtracted from the original image. This process was carried out separately for each frame, since the pattern shifted from image to image.

\medskip

\subsection{Frame Alignment}
Difference imaging requires precise frame alignment in order to produce a proper subtraction. Since CSTAR had fixed pointing towards the SCP it was necessary to rotate each frame to the same rotation angle as the reference frame. We initially used the difference between the date of each target image and the reference frame to calculate the angle offset. We rotated each frame using a cubic convolution interpolation that approximates the optimum interpolation function, as implemented in the IDL function ROT. This function can optionally apply a translation to the target image.

As seen in previous reductions of CSTAR data, we found that the time stamps generated by the local computer drifted as the season progressed, leading to improper alignment of the frames and poor image differencing. This drift was expected due to the lack of network time synchronization at Dome A. Furthermore, we noticed that the location of the SCP (determined from the astrometrically-calibrated master frame of \citealt{Wang2013}) moved across the detector as a function of time. While the exact nature of this motion has yet to be determined, it is hypothesized to be due to the drift of the Antarctic ice shelf, the effect of winds, heating due to changes in solar elevation, or a combination of these.

\begin{figure}[t!]
\centering
\includegraphics[width=70mm]{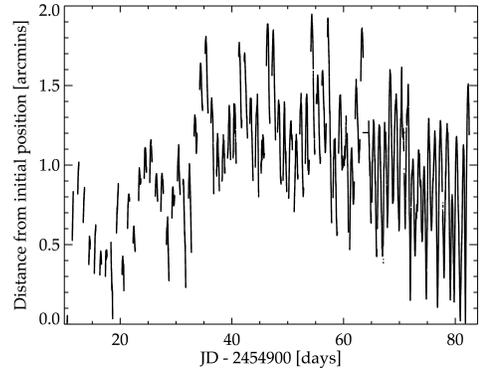}
\caption{The movement if the SCP from its initial position in \textit{g} as a function of Julian date. There are variations in the movement of the SCP both on a daily level and with a period of 28.8 days.\label{fig:scp}}
\end{figure}

We solved both issues by carrying out aperture photometry on all images and matching star lists using the DAOPHOT, DAOMATCH and DAOMASTER programs \citep{Stetson1987}. Once the proper rotation and translation values were determined for each image, they were applied using the function described above. Figure~\ref{fig:scp} shows the displacement of the SCP from its initial position as a function of time for the \textit{g} images. Using a Lomb-Scargle periodogram \citep{Lomb, Scargle}, we found 2 significant periods at 1 and 28.8 days; these two periods are likely due to changes in solar elevation \& lunar tides, respectively.

\medskip

\medskip

\section{Photometry}

Previous analysis of CSTAR data \citep{Zou2010, Zhou2010, Wang2011, Wang2013} were based on aperture photometry, which involves the placement of apertures of fixed radii on the centroid of a target star, summing the enclosed flux, and correcting for background light estimated from an annular region. Aperture photometry works best when the ratio of PSF size to stellar separation is well below 1; in other words, when blending and crowding are not significant. The 2009 CSTAR data did not meet these standards, making the use of aperture photometry undesirable. Instead, we used difference image analysis (hereafter, DIA) to measure changes in stellar flux between each science frame and a reference frame. DIA has been shown to work well in crowded fields with data from small aperture telescopes \citep{Pepper2007}. Our code is a version of the optimal frame subtraction routine \textit{ISIS} \citep{AlardLupton}. It is written in C and requires 45 CPU seconds to difference a single frame using Intel Quad-Core Xeon 2.33GHz/2.8GHz processors. 

\medskip

\subsection{Reference Frame\label{sc:ref}}
A key component to DIA is the generation of a high-quality reference frame. Typically this frame is generated by median-combining many individual frames with high $S/N$ and the best seeing, obtained throughout the observing season. This approach is not feasible for our images due to the spatial variations in the defocused PSFs and the continuously changing orientation of the field. When we carried out this process we found the PSF of the reference frame became broader and more Gaussian-like in shape compared to the torus-like PSF of individual images, making it more difficult to convolve and subtract. Therefore we selected a single frame obtained near the end of the observing season (free of satellite trails, clouds or other undesirable features) as the reference for each band. Subsections of the reference frames are shown in Figure~\ref{fig:ref}. Unfortunately, unlike using a median combination of images, using a single image as a reference frame does not minimize the possible noise in the reference. However, because the frame was selected to be of better or equal quality than the science frames, we expect the photometric uncertainty will not increase by more than $\sim\sqrt{2}$.

\medskip

\subsection{Kernel definition}

Typically, DIA routines use an adaptive kernel, $K(x,y)$, defined as the combination of 2 or more Gaussians. While effective at modeling well defined, circular PSFs, this kernel has difficulty properly fitting other PSF shapes. Therefore, we used a Dirac-$\delta$ function kernel to compensate for our non-circular, irregular PSF shape. We redefined the kernel as

\begin{equation}
	K(x,y) =  \sum_{\alpha=-w}^w\sum_{\beta=-w}^w c_{\alpha,\beta}(x,y)K_{\alpha, \beta}(u,v)
\end{equation}

\noindent{where $K_{\alpha, \beta}$ is a combination of $(2w+1)^2$ delta function basis vectors and $K_{0,0}$ is the centered delta function \citep{Miller2008}. We redefined our basis vectors to ensure a constant photometric flux ratio between images \citep{Alard2000, Miller2008}. In the case of $\alpha \neq 0$ and $\beta \neq 0$,}

\begin{equation}
	K_{\alpha, \beta}(u,v) = \delta(u-\alpha,v-\beta) - \delta(u,v) 
\end{equation}

\noindent{while for $\alpha = 0$ and $\beta = 0$,}
\begin{equation}
	K_{0,0}(u,v) = \delta(u,v).
\end{equation}

\begin{figure}[t!]
\centering
\includegraphics[width=90mm]{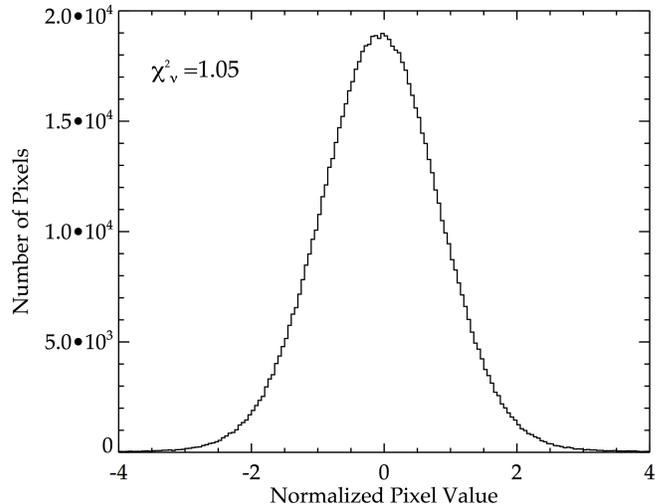}
\caption{The normalized distribution of pixel values for an average differenced frame. We excluded pixels near stars close to saturation or near to the edge of the frame. A proper subtraction should yield $\chi^2_{\nu} \sim 1$; this frame has $\chi^2_{\nu} = 1.05$. \label{fig:hist}}
\end{figure}

\begin{figure}[b!]
\centering
\includegraphics[width=67mm]{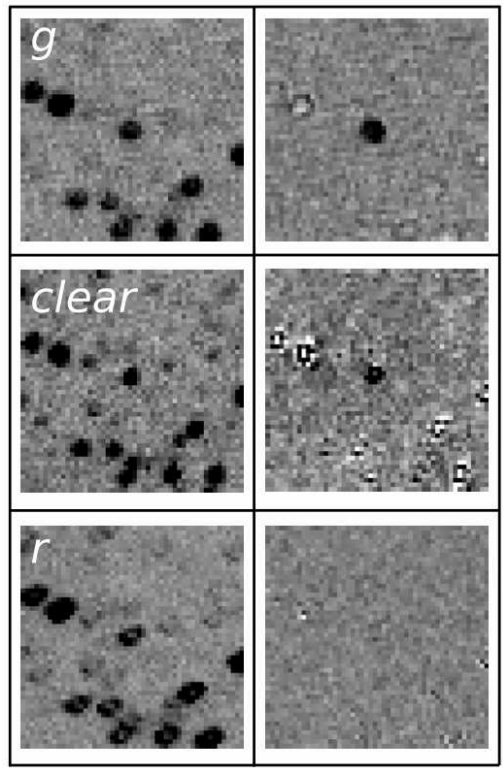}
\caption{$50\times50$~pixel subsections of each reference frame (left) and 3 differenced science frames (right). Each subsection is centered on the recovered RR Lyrae \#n058002 in \textit{g} (top), {\cl} (middle) and \textit{r} (bottom). The right frames show differenced frames at varying points of the RR Lyrae phase. \textit{g} shows the star at maximum, {\cl} shows the star at mid brightness and \textit{r} shows the star at minimum. Coincidentally the reference frame was at the minimum of the RR Lyrae phase and thus no change appears in the last differenced frame instead of a negative value. Systematics can be seen in the {\cl} differenced frame due to the lens frosting. Each frame has a pixel scale of $15\arcsec$/pix. The color scale has been inverted for clarity, with darker colors denoting a positive residual. \label{fig:diff}}
\end{figure}

Stamps were taken around bright, isolated stars to solve for the coefficients $c_{\alpha, \beta}(x,y)$ using the least-squares method. Since the PSF in our images was spatially varying, we applied a $5\times5$, $1^{\rm st}$-order adaptive kernel across the frames. We found this kernel size minimized the $\chi^2_{\nu}$ solution \textit{without} over-fitting the data as shown in Figure~\ref{fig:hist}. We also allowed $c_{0,0}(x,y)$ to be spatially variable to compensate for imperfect flat field corrections. The typical quality of the differencing for each band in shown in Figure~\ref{fig:diff}.  

\medskip

\medskip

\subsection{Flux Extraction}
We extracted the differential fluxes using the IDL version of the DAOPHOT package. The highly irregular PSF shape caused trouble for many stellar detection algorithms such as DAOPHOT's FIND, which would detect a single source multiple times in neighboring pixels while skipping other sources entirely. Therefore, we visually identified a total of 2086 sources in our \textit{g} reference frame that were always contained in the field of view and we included the transformed coordinates of the 165 variable stars previously detected by \citet{Wang2011,Wang2013, WangS2014}. The area of continuous coverage in our reference frames spans a 450 pixel ($1.875^{\circ}$) radius circle centered on the SCP. 

We set the photometry aperture at a 5 pixel radius ($1.25\arcmin$) with a sky annulus spanning 8 to 10 pixels ($2-2.5\arcmin$). The differential flux was then combined with the flux from the reference frame and corrected for exposure time. We used the astrometric data from the 2008 and 2010 master frames of \citet{Wang2011,Wang2013} to convert our reference frame $(x,y)$ coordinates to celestial ones. 

We applied an exposure time correction based on the given exposure times in the image header. However, this did not properly scale the flux for \textit{r} exposures taken between JD 2454910 and 2454955. We determined the effective exposure times for these images as follows. We measured the peak magnitudes in successive cycles of previously-known contact binaries and RR Lyraes (since their variations are known to be highly stable) and compared those values to the ones measured on the day when our reference images were obtained. We determined that images taken between JD 2454910 and 2454935 had exposure times $5\times$ shorter than expected, while those obtained between JD 2454935 and 2454955 had exposure times $4\times$ shorter than expected. We hypothesize that the telescope was commanded to expose for 5 or 20 seconds during the respective time intervals but actually exposed for 1 and 5 seconds, respectively. We therefore adopted the latter exposure times in our analysis. We scaled the DAOPHOT errors to match what was expected based on the new exposure times and the error after JD 2454955 to ensure they would not be underestimated. Finally, we applied an absolute photometric calibration based on the synthetic $griz$ magnitudes of Tycho stars from \citet{Ofek2008} following a similar method to previous reductions \citep{Wang2011, Wang2013}. The {\cl} photometry was calibrated to the Tycho $V_T$ system \citep{Grossmann1995}. 

\medskip

\subsection{Trend Removal}
We did not remove any frames based on their quality (sky background, clouds, etc.) except for frames with obvious filter frosting in {\cl}. All data points in each light curve were included regardless of their ``outlier'' status and we relied on fully propagated errors to provide statistical significance to any deviations. We used the ensemble photometry to identify and remove systematic trends due to instrumental or processing effects that were present in multiple light curves.

These systematics could have several origins. The first is due to the movement of the stars around the detector. As the stars move across the detector they may experience slight fluctuations in their light due to inconsistencies on the detector or our flat field. Convolving the reference frame with a kernel matrix could also introduce systematics if the kernel was improperly solved. Finally lens vignetting, subtle changes in the airmass across the wide field and filter frosting may create non-astrophysical fluctuations in the light curves. We used the Trend Fitting Algorithm (TFA) \citep{Kovacs2005} as implemented in the VARTOOLS package \citep{Hartman2008} to compensate for these systematics. We used 150 stars spanning a wide range of fluxes and locations in the frame that did not exhibit any discernible variations of an astrophysical nature (e.g., eclipsing binaries, periodic variables, etc.) as templates for the trend removal.

We also implemented an alternative trend-removal procedure to deal with the unusual systematics that may arise from the unique nature of the CSTAR observations, namely stars that describe a daily circular motion across the FOV. We ran a Lomb-Scargle \citep{Lomb, Scargle} period search on every light curve and identified its most significant period within $0.98 < P < 1.02$~d. We selected the equivalent of TFA reference stars with the same period as the target object, light curve \textit{r.m.s.}$<0.2$~mag, and located within 110 pixels in radius of the object being considered. Each reference star was scaled to match the amplitude of the variation in the target star. All of the resulting scaled reference light curves were median combined and subtracted from the target light curve. Stars known to exhibit periodic variation of astrophysical origin had this signal removed (a process commonly referred to as pre-whitening) before conducting this correction. 

We calculated the \textit{r.m.s.} on 30-minute timescales for the light curves corrected with TFA and the alternative approach and selected the one that exhibited the lowest dispersion.

\medskip
\section{Noise}
Small aperture telescopes may exhibit systematic effects that are not always present in their larger counterparts, especially when considering the effects of a defocused PSF. When a star is isolated and free from the effects of crowding and blending, purposefully defocusing a telescope has been shown to greatly decrease the photometric dispersion by minimizing flat-field uncertainties \citep{Southworth2013}. The unanticipated defocusing of the CSTAR system presented difficulty as the telescope was not designed for such techniques due to its crowded field and large pixel scale. Adequate understanding of noise is key to understanding the difference between true signals and systematics.

\subsection{Poisson Deviation}
The noise in a differenced frame comes from two sources: the science frame and the convolved reference frame. \citet{AlardLupton} model the effects of noise in a differenced frame building on the typical assumption of $\sigma = \sqrt{I_N}$, where $I_N$ is the photon counts in the frame. The \textit{Poisson deviation} is defined as $\delta = \sqrt{I_N+R_N \otimes K^2}$ to describe the effects of the noise in each differenced frame, with $R_{N}$  being the photon counts from the reference \citep{AlardLupton, Alard2000}. 

Normalizing the pixel values in the differenced frame by $\delta$ is a good way to determine if noise is being added by the difference imaging routine. The normalized points should show a Gaussian-like distribution around 0 with a standard deviation close to 1. Figure~\ref{fig:hist} shows the histogram of pixel values for a typical differenced frame normalized by $\delta$. The mean of the residuals is $\sim-0.05$ with a standard deviation of $\sim1.13$. We calculated the $\chi^2_{\nu}$ of this differenced frame to be $\sim1.05$. 

\begin{figure}[t!]
\center
\includegraphics[width=85mm]{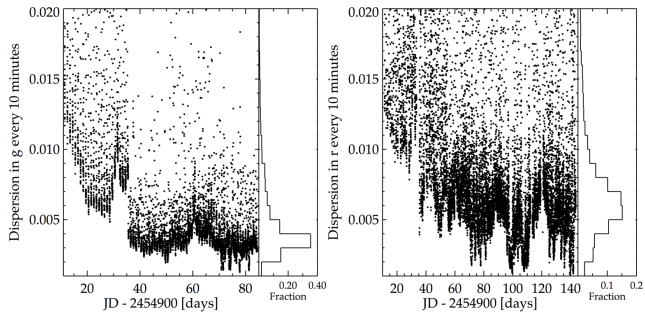}
\caption{The change in photometric precision for 13 bright, non-saturated stars with magnitude $< 8$ in \textit{g} $\&$ \textit{r} as a function of Julian date on 10 minute intervals.\label{fig:chdisp}}
\end{figure}

\begin{figure}[h!]
\centering
\includegraphics[width=85mm]{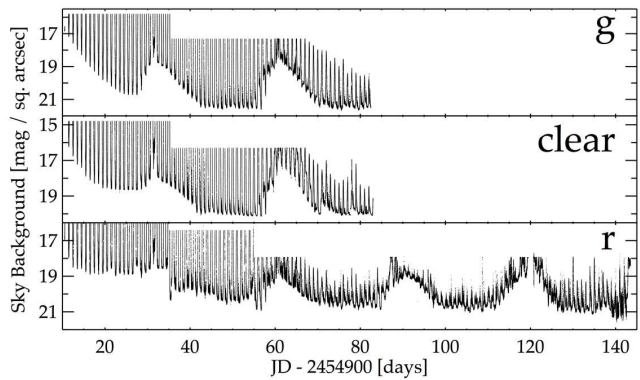}
\caption{Sky background in magnitudes per sq. arc second  vs Julian date for \textit{g} (top), {\cl} (middle) and \textit{r} (bottom). The lunar cycle can be seen at JD 2454900 $+\sim$ 30, 60, 90 and 120. The step-like change in the uppermost values of the background is due to a change in exposure time from 5 to 20 seconds.\label{fig:back}}
\end{figure}

\begin{figure}[ht!]
\centering
\includegraphics[width=85mm]{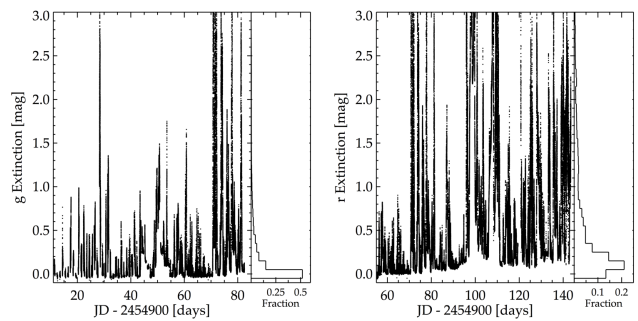}
\caption{Differential extinction in magnitude vs Julian date for \textit{g} (left) and \textit{r} (right) based on aperture photometry of the $<1000$ brightest stars with $<0.2$~mag photometric error in each band. \label{fig:extinct}}
\end{figure}

\subsection{Noise Model}

Convinced our routine was not adding additional systematics to each frame we then developed a model for the noise. Astronomical noise from brighter sources is dominated by the number of photons from the source and increases as $\sqrt{I_N}$. This model is the expected dispersion for stars with high photon counts relative to the sky background.

Figure~\ref{fig:chdisp} shows the change in light curve dispersion with time for stars with \textit{g} $\&$ \textit{r} $< 8$. Monitoring the change in light curve dispersion over long time intervals ($\sim$ 10 minutes) allows us to determine the observing conditions from the site. DIA uses least squares fitting to match the PSF changes and photometric zeropoint offsets between the reference and science frames caused by clouds and/or changes in airmass. These effects decrease the photon count above the background and increase the photometric dispersion with time. In \textit{g} we find bright stars have a dispersion level of $<0.01$ mag for $90\%$ of the observing season and $<0.005$ mag for $66\%$ of the observing season in 10 minute intervals. In \textit{r} we find bright stars have a dispersion level of $<0.01$ mag for $72\%$ of the observing season and $<0.005$ mag for $23\%$ of the observing season.

The noise for fainter objects is dominated by the sky background, which is present in all 2009 CSTAR images as shown in Figure~\ref{fig:back}. Clear modulation in the levels of the minimum sky background can be seen at a period of 28.8 days due to contamination from the Moon. The step-like nature of the sky background marks the delineation between exposure times of 5 and 20 seconds near JD 2454935. 

The noise from the sky is usually modeled as $\pi r^2 I_{\rm sky} $, where r is the pixel radius of the aperture and $I_{\rm sky}$ is the photon counts from the sky in an individual pixel. We can create a complete model for the expected noise in each differenced frame as shown in equation~\ref{eq:noise}. The factor of 2 included in the model takes into account the additional noise introduced by using a single image for the reference frame instead of a median-combined set of images, due to the reasons described in \S\ref{sc:ref}. We then have

\begin{equation}
\sigma = \sqrt{2[I_N + \pi r^2 (I_{sky})]}
\label{eq:noise}
\end{equation}

\subsection{Extinction from Dome A}
One of CSTAR's primary objectives was to determine the observing conditions at Dome A for a possible permanent installation. The photometric offset between frames, also known as differential extinction, can act as a proxy for the amount of cloud cover during an observing season. The analysis of previous CSTAR observations found the site to have \textit{i}-band extinction due to clouds less than 0.4~mag for $70\%$ ($80\%$) of the dark time and less than 0.1~mag for $40\%$ ($50\%$) of the dark time in 2010 (2008) \citep{Wang2013, Wang2011}.

We measured the extinction between our \textit{g} and \textit{r} reference frames and each science frame using aperture photometry with DAOPHOT \citep{Stetson1987}. We selected $<1000$ of the brightest stars with photometric error $<0.2$~mag within 400 pixels of the center of each frame and then used DAOMATCH/DAOMASTER to find an initial magnitude offset. We calculated the mean photometric offset for each frame using an error-weighted sigma clipping technique and found the typical uncertainty in the mean offset to be $\sim$0.015~mag.

Comparing results from both \textit{g} and \textit{r}, when both telescopes were operational, we find very good agreement with a difference in the extinction values of $\langle \Delta(g-r) \rangle = 0.02\pm0.01$~mag. We find the extinction due to clouds at Dome A was less than 0.4~mag for $63\%$ of the dark time and less than 0.1~mag for $40\%$ of the dark time. Figure~\ref{fig:extinct} shows a time-series and histogram of extinction values. The extinction in \textit{r} was calculated using the data obtained after JD 2454955, to avoid possible biases due to the issues described in \S4.3. We believe the $\sim0.1$ mag mean increase in extinction after JD 2454993 is likely due to minimal filter frosting as we find the average number of stars per exposure drops by $\sim 5\%$ after JD 2454993. We excluded {\cl} from this analysis because of the rampant filter frosting which would be indistinguishable from extinction due to clouds.

\clearpage

\subsection{Scintillation Noise}
The photometry of bright stars in our sample is constrained by the scintillation limit of the telescope, so we added this feature to our noise model. \citet{Young1967} modeled the effect of scintillation as a function of telescope diameter ($d$, in cm), altitude ($h$, in m), airmass ($X$) and exposure time ($t_{\rm ex}$, in s).  We adopt an updated version of this model by \citet{Hartman2005}:

\begin{equation}
S = S_0 d^{-\frac{2}{3}}X^{\frac{7}{4}}e^{-h/8000}(2t_{\rm ex})^{-\frac{1}{2}}
\label{eq:scin}
\end{equation}

\noindent{where $S_0 \sim 0.1$ \citep{Young1967,Hartman2005}. Given the wide field of view of CSTAR, the airmass values of stars in our images are $1.01\pm0.005$. The effective elevation of Dome A is $h=5100$~m, taking into account the reduced pressure (560mb) due to its polar location. Given a typical exposure time of 20s, we find the scintillation limit to be $1.2 \pm 0.1$~mmag (with the variation due to the airmass range being considered).}

\begin{figure}[t!]
\center
\includegraphics[width=85mm]{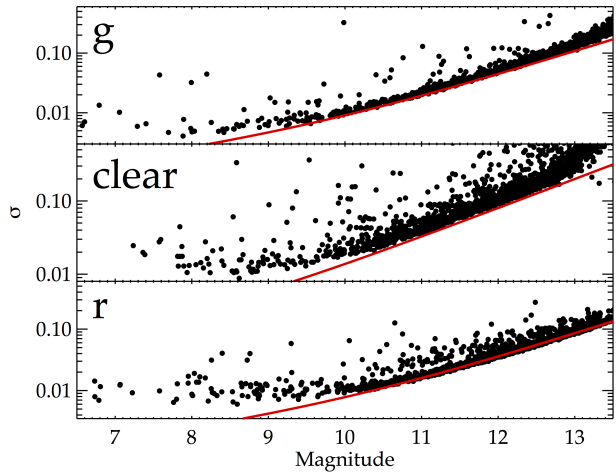}
\caption{Photometric precision in \textit{g} (top), {\cl} (middle) and \textit{r} (bottom) for the differencing analysis of the 2009 CSTAR data during the entire observing season. The sample consists of 2086 sources with complete seasonal coverage, fully contained in the FOV and 162, 164, 165 previously known variable stars. The red lines are simple models for the expected error as a function of magnitude in each band (equation 4) with a scintillation floor of 1.2 mmag. At \textit{g} $\&$ \textit{r} we reach within a factor of $\sim$ 3 of the expected scintillation noise at a magnitude of 8. The increased dispersion at {\cl} $<$10 is likely due to intermittent frosting of the filter due to power failures.\label{fig:disp}}
\end{figure}

Our final model for the noise and the observed photometric precision in each band is plotted in Figure~\ref{fig:disp}. The lowest dispersions were found for stars in the magnitude range of $\sim 7 - 10$ in all three bands, where we reached within a factor of 3 of the scintillation noise in \textit{g} \& \textit{r}. The additional dispersion at $m<10$~mag in the {\cl} data is most likely due to partial and intermittent frosting of the filter.

\medskip

\medskip

\section{Variable Stars in the CSTAR Field}

One of the main goals of this study was to confirm the ability of the DIA code to detect stellar variability even in the hostile photometric environment of the crowded and defocused images of the 2009 CSTAR data. Our search for variability included a subset of stars with complete seasonal coverage (2086 in \textit{g}, 2080 in {\cl} and 2086 in \textit{r}). We also included previously-detected variable stars located in our reference frames from \citet{Wang2011,Wang2013} (158 in \textit{g}, 158 in {\cl} and 159 in \textit{r}) and transiting exoplanet candidates from \citet{WangS2014} (2 in \textit{g}, 3 in {\cl} and 3 in \textit{r}) for a total of 2246 stars in \textit{g}, 2241 stars in {\cl} and 2248 stars in \textit{r}.

We applied the following searches to the \textit{g} and \textit{r} data only. We used the {\cl} search results \textit{only} as a confirmation of variability in either \textit{g} or \textit{r}, given the much shorter span of the {\cl} data and the impact of intermittent frosting on its filter. The {\cl} panels in Figures~\ref{fig:rms}-\ref{fig:perhist} are only shown for completeness. 

\subsection{Search for Variability \label{subsec:varmetric}}

\begin{figure}[t!]
\center
\includegraphics[width=85mm]{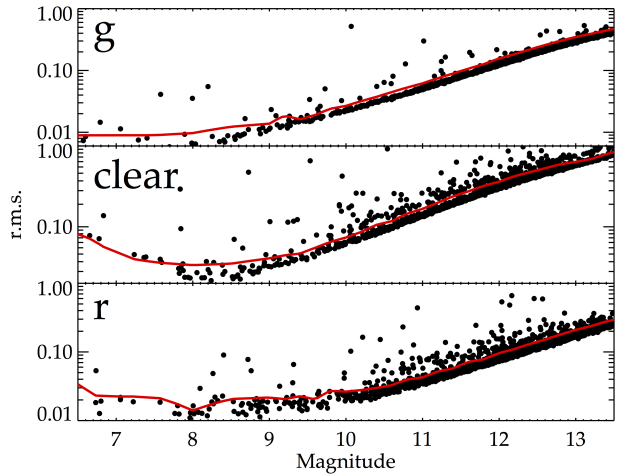}
\caption{\textit{r.m.s} variability statistic for \textit{g} (top), {\cl} (middle) and \textit{r} (bottom). The red line denotes the upper $2\sigma$ cutoff for a variable candidate. We identified 76 candidates in \textit{g},  285 in {\cl} and 270 in \textit{r} using this test.\label{fig:rms}}
\end{figure}

\begin{figure}[t!]
\center
\includegraphics[width=85mm]{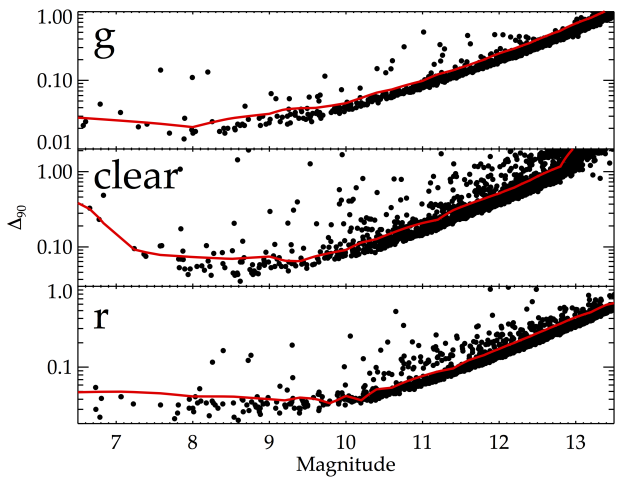}
\caption{$\Delta_{90}$ variability statistic for \textit{g} (top), {\cl} (middle) and \textit{r} (bottom). The red line denotes the upper $2\sigma$ cutoff for a variable candidate. We identified 111 candidates in \textit{g},  420 in {\cl} and 351 in \textit{r} using this test.\label{fig:d90}}
\end{figure}

We employed a combination of 3 variability metrics, following the approach of \citet{Wang2013}. First we computed the \textit{r.m.s.} of all stars and the upper 2$\sigma$ envelope as a function of magnitude, as shown in Figure~\ref{fig:rms}; objects lying above this limit are likely to be genuine astrophysical variables. Next, we computed the magnitude range spanned by 90\% of the data points of every light curve (hereafter, $\Delta_{90}$) and its upper 2$\sigma$ envelope as a function of magnitude; the results are plotted in Figure~\ref{fig:d90}. Since we wished that both statistics be based on ``constant'' stars only and not be biased by large-amplitude variables, both envelopes were calculated in an iterative fashion. We discarded objects located above the median by more than the difference between the median value and the minimum value.

\begin{figure}[t!]
\center
\includegraphics[width=85mm]{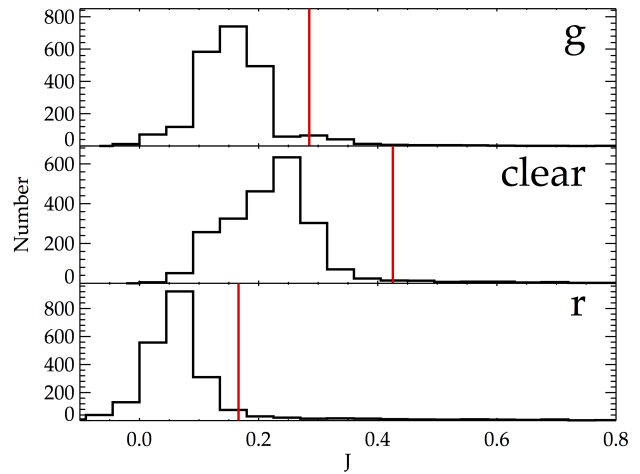}
\caption{\textit{J} variability statistic for \textit{g} (top), {\cl} (middle) and \textit{r} (bottom). The red line denotes the $3\sigma$ cutoff for a variable candidate. We identified 147 candidates in \textit{g}, 98 in {\cl} and 222 in \textit{r} using this test.\label{fig:jstet}}
\end{figure}

Finally, we computed the Welch-Stetson \textit{J} variability statistic \citep{Stetson1996} including the necessary rescaling of DAOPHOT errors. The \textit{J} statistic is useful to detect variability during short time spans, such as the 5 and 20 second sampling of the CSTAR data, since it computes the significance of photometric variability between two adjacent data points. The \textit{J} statistic is expected to produce a distribution of values with a mean value close to zero for the ``constant'' stars and a one-sided tail towards positive values for the ``variable'' stars. We considered objects lying above the $+3\sigma$ value as variable. The results of this statistic are plotted in Figure~\ref{fig:jstet}. 

We considered a star to be variable if the star passed all 3 of the above tests in either \textit{g} or \textit{r} and passed 2 or more tests in at least 1 of the remaining 2 bands. We rejected any star that was identified as a candidate variable but had a primary Lomb-Scargle period \citep{Lomb, Scargle} between 0.98 and 1.02 days with $S/N>100$. We interpret these objects as being biased by aliased systematics as CSTAR greatly suffers from aliases of 1 day. If these stars later showed statistically significant non-aliased periods, we allowed them into the periodic sample described in \S~\ref{subsec:permetric}. 

\subsection{Search for Periodicity \label{subsec:permetric}}

\begin{figure}[t!]
\centering
\includegraphics[width=80mm]{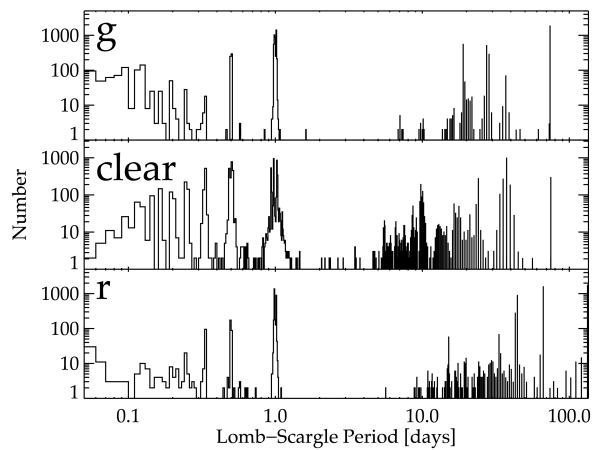}
\caption{Results of the Lomb-Scargle period search. The top 5 periods were searched for between 0.01 and 74 days in \textit{g} and {\cl} or 135 days in \textit{r}. We identified 28 periodic candidates in \textit{g}, 10 in {\cl} and 61 in \textit{r} in our periodicity search. \label{fig:perhist}}
\end{figure}

The three metrics described above are sensitive to variable stars with statistically large amplitude variations compared to other stars of similar magnitude; unfortunately these tests lack the ability to detect small amplitude, periodic variations. To compensate for this we ran a search for periodicity based on the Lomb-Scargle (LS) method \citep{Lomb, Scargle} as implemented by \citet{Wang2011}. We computed the 5 highest signal-to-noise periods of each star in our sample between 0.01 and 74 days in \textit{g} and 135 days in \textit{r}. We binned these periods into bins of 0.01 days and discarded periods with a count of 10 or more stars as impostor periods. Figure~\ref{fig:perhist} shows the result of this analysis.  

We also ran the Box Least Squares algorithm (hereafter BLS) to search for eclipse-like events which may have eluded our previous variability searches \citep{Kovacs2005}. BLS looks for signals characterized between two discrete levels, the transit (high-level) and the occultation (low-level). We searched each light curve for transits with a range of 0.01 and 0.1 of the primary periods between 0.1 and 74 days in \textit{g} and 135 days in \textit{r}. We allowed for 10,000 trial periods and 200 phase bins.

We only considered a star to be periodic if the period had a $S/N > 700$ and could be recovered in at least 2 of the 3 bands or with the BLS search. We removed any star from our total variable sample if it was located within 10 pixels of a brighter candidate variable star as the differenced residuals were likely biasing the flux measurements within the 5 pixel aperture. 

\section{Results}

Table~\ref{tb:varmetric} summarizes the results of the variability search while Table~\ref{tb:permetric} summarizes the results of the periodicity search. Using these complementary techniques, we identified a total of 46 \& 92 variables in \textit{g} and \textit{r}, respectively. Taking into account objects that were identified in both bands, our final catalog contains 105 objects. 37 of these objects were not identified as variables in previous CSTAR papers. Table~\ref{tb:rec} lists the properties of these stars.

The defocused nature of the observations likely aided in the identification of 7 stars as variable. These stars were either close to or fully saturated in the \textit{i} data from 2008 and 2010. The American Association of Variable Star Observers (AAVSO) has previoulsy catalogued 4 of these stars. The AAVSO has classified 2 of these variables, \#p09-004 and \#p09-002, as a slowly-varying and a non-periodic semi-regular variable, respectively. After our LS search we found both of these stars to have periods passing our threshold criterion of 60.5~d and 22.2~d respectively. \#p09-003 is archived in the  AAVSO database as a miscellaneous variable with a period of 73.3~d. We found this variable to be semi-regular, currently exhibiting a main period of 16.6~d. We have also recovered the variability in \#p09-007, which is classified as a $\delta$ Scuti star with a period of 0.12~d. 

\begin{deluxetable}{lccccccc}
\tablewidth{0pt}
\tablecaption{Number of Stars Passing each Variability Metric \label{tb:varmetric}}
\tablehead{\colhead{Band} & \colhead {$J$} & \colhead \textit {r.m.s.} & \colhead{$\Delta_{90}$} & \colhead{All} & \colhead{Aliases} &\colhead{Proximity} & \colhead{Final}}
\startdata
\textit{g} 	& 147 & 76 & 111 & 35 & 0 & 10 & 25 \\
{\cl}        	& 98	 & 285 & 420 & 50 & 2 & 15 & 33 \\ 
\textit{r} 	& 222 & 270 & 351 & 69 & 2 & 14 & 53
\enddata
\end{deluxetable}

\begin{deluxetable}{lccccc}
\tablewidth{0pt}
\tablecaption{Number of Stars Exhibiting Significant Periodicities \label{tb:permetric}}
\tablehead{\colhead{Band} & \colhead{Variables}                & \multicolumn{2}{c}{Periodicity Search} & \colhead{Total}\\ 
                          & \colhead {from \S~\ref{subsec:varmetric}} & \colhead {LS} & \colhead {BLS} & }
\startdata
\textit{g} & 7 & 11 & 17 & 28 \\
{\cl}        & 4 & 2 & 8 & 10 \\
\textit{r} & 22 & 43 & 18 & 61
\enddata
\end{deluxetable}
\begin{figure}[t!]
\centering
\includegraphics[width=85mm]{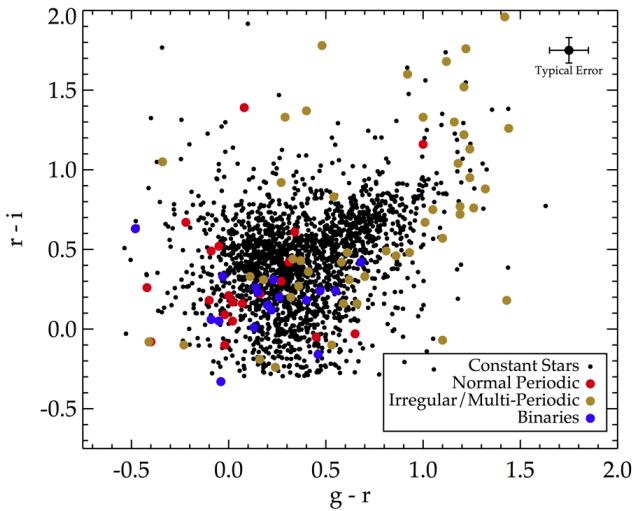}
\caption{Color-color diagram for stars in our 2009 CSTAR sample with three-band photometry (the i-band data is from the 2010 CSTAR photometry of \citet{Wang2013}). Small black points denote the data points the constant stars. Red points are regular periodic variables such as RR Lyrae, $\delta$~Scuti and $\gamma$~Doradus. Gold points are irregular, multi-periodic and long term variable stars. Blue points are the eclipsing binaries.\label{fig:ccplot}}
\end{figure}

A major advantage of the 2009 CSTAR dataset is the addition of 3 color photometry. Variable stars in the CSTAR field now have the unique opportunity to be studied for variations in both time and color. Figure~\ref{fig:ccplot} is a color - color diagram for stars in our sample with \textit{gri} magnitudes. We find $\sim91\%$ of the stars in our sample have \textit{g}$-$\textit{r} $>0$. This is consistent with the CSTAR field being directed towards the galactic halo and confirms previous and current variable star searches in the field finding many irregular and multi-periodic RGB or AGB like stars. Indeed we find normal pulsators, such as RR Lyraes or $\delta$ Scuti stars, multi-periodic and irregular variables have $\langle g\!-\!r \rangle \sim 0.59$. In contrast the eclipsing binaries, which are expected to have a wide variety of ages along the main sequence, have $\langle g\!-\!r \rangle \sim 0.22$. 

\begin{figure*}[ht!]
\centering
\includegraphics[width=120mm]{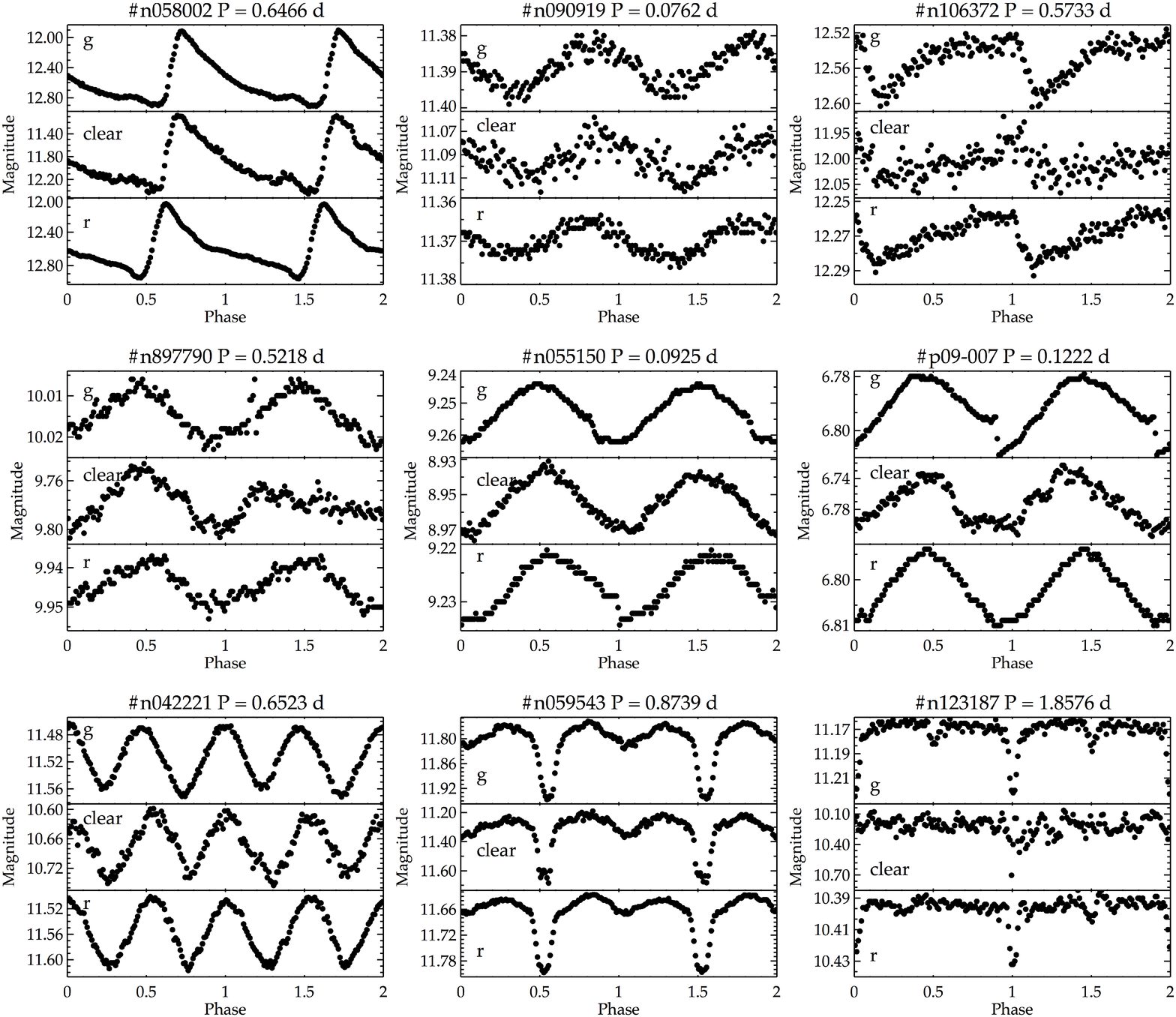}
\caption{Light curves for 9 variable stars in \textit{g} (top), {\cl} (middle) and \textit{r} (bottom) showcasing the different types of objects present in our sample (from top left to bottom right): RR Lyrae (\#n058002); periodic variable (\#n090919); periodic variable (\#n106372); $\gamma$~Doradus (\#n897790); periodic variable (\#n055150); $\delta$~Scuti (\#p09-007);  contact binary (\#n042221); semi-detached binary (\#n059543); and detached binary (\#n123187). The light curves have been phased and binned into 200 data points. \label{fig:recover}}
\end{figure*}

Figure~\ref{fig:recover} shows the light curves of 9 variable stars in our data. \#p09-007 is an example of a bright star ($g\sim r \sim 6.8$~mag) that was saturated in CSTAR observations carried out during other winter seasons when the array was in focus. The defocused nature of our images allowed it to remain below the saturation limit, enabling a period determination of 0.122~d. Previous studies of \#n106372 classified this star with a period of 12.5~d \citep{Wang2011}. We recover the star as periodic but with a significant period of 0.57~d in all bands. The remaining variables shown in the Figure were present in the 2008 and 2010 data sets and span a variety of types and periods.

\begin{figure*}[ht!]
\center
\includegraphics[width=150mm]{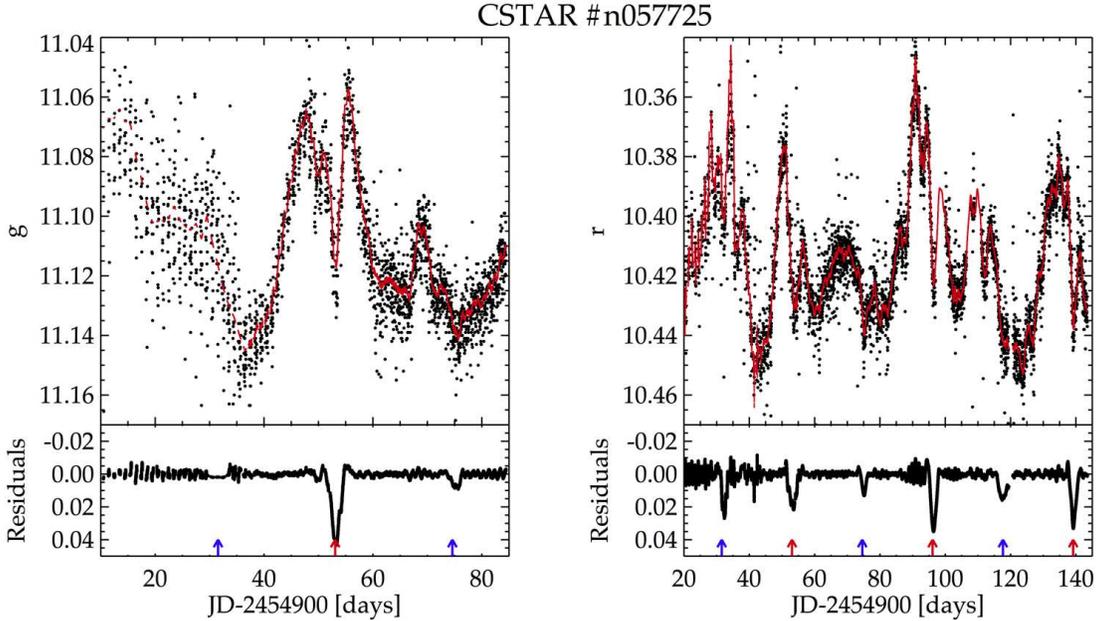}
\caption{Light curves of  \#n057725, a likely Population II Cepheid in an eclipsing binary system showing complex variability. The top panels show the 2009 \textit{g} and \textit{r} light curves with the smoothed light curve over-plotted. The bottom panels highlight the eclipse-like events that take place every 43.2~d. Red arrows mark the expected time of primary eclipse and blue arrows mark the expected time of secondary eclipse. \label{fig:ceph}}
\end{figure*}

Figure~\ref{fig:ceph} shows the \textit{g} and \textit{r} light curves of \#n057725. This variable exhibited very regular, Cepheid-like pulsations in the 2008 \textit{i} data and a much more complex light curve structure in the 2010 \textit{i} data, with clear evidence of eclipses. The 2009 light curves show enhanced variability in the cepheid-like modulation of the light curve. The expected times of eclipse are highlighted with red arrows for primary eclipses and blue arrows for secondary eclipses. We applied a smoothing kernel to the light curves to aid in the recovery of the suspected binary eclipses. We find we recover both the primary and secondary eclipses in \textit{g} $\&$ \textit{r} at the expected eclipse times.

\section{Conclusions and Summary}

We have presented a technique useful for the reduction of crowded, defocused data. The 2009 Antarctic winter season observations by CSTAR at Dome A suffered from intermittent filter frosting, premature power failures and a defocused PSF. Even with these technical issues the system obtained a total of $\sim 10^6$ scientifically-useful images in the 3 operating bands.

Each frame underwent extensive pre-processing including bias subtraction, flat fielding, background subtraction, electronic fringe subtraction and frame alignment. We used a combination of difference imaging with a delta function kernel and aperture photometry to compensate for the highly crowded, blending and defocused frames. We applied the Trend Fitting Algorithm and an alternative de-aliasing trend removal technique to correct for systematics resulting from detector variations or improper kernel fits. 

We applied 3 variability tests, one periodicity search and one transit search to all light curves. We recovered 68 of 165 previously-known variable stars within our magnitude limit ($g\sim r \sim 13.5$~mag) and identified 37 previously undiscovered variables in CSTAR data sets.

We plan to use this image-processing technique in the near future to search for astrophysical transients in the 2010 CSTAR \textit{i} data, and to analyze ongoing observations with a similar system operating from the Bosque Alegre Astrophysical Station in C\'ordoba, Argentina. The differencing code is freely available upon request to the corresponding author (RJO).

\clearpage

\acknowledgments
RJO \& LMM acknowledge support from the George P. and Cynthia Woods Mitchell Institute for Fundamental Physics and Astronomy; the Mitchell-Heep-Munnerlyn Endowed Career Enhancement Professorship in Physics or Astronomy; and by the Joint Center for the Analysis of Variable Star Data, funded by the Indo-US Science and Technology Forum. 

RJO acknowledges support of the National Science Foundation Grant No.~1209765 for the 2012 East Asian and Pacific Summer Institute Fellowship, during which the difference imaging code was written. Thanks are due to Dr.~Lingzhi Wang for explaining the reduction procedures for other CSTAR datasets and granting access to the corresponding master frames, and to Dr.~Ji-Lin Zhou for serving as the host at Nanjing University for the 2012 EAPSI Fellowship. 

This project is supported by the Commonwealth of Australia under the Australia-China Science and Research Fund, the Australian Research Council and the Australian Antarctic Division.

\bibliographystyle{apj}
\bibliography{references}

\clearpage 

\clearpage

\LongTables
\begin{deluxetable*}{llrrrrrrrrc}
\tabletypesize{\tiny}
\tablewidth{0pt}
\tablecaption{Variable stars detected in the 2009 CSTAR observations \label{tb:rec}}
\tablehead{\colhead{CSTAR ID} & \colhead {GSC} & \colhead {R.A.} & \colhead{Dec.} & \colhead{\textit{i}} & \colhead{\textit{g}} & \colhead {\cl} & \colhead {\textit{r}} & \colhead{Type} & \colhead{P (d)} & \colhead{Catalog}\tablenotemark{1}} 
\startdata
n020508 & S74D000440 & 13:01:58.40 & -87:39:56.30 &   8.99 &   9.53 &   9.01 &   9.30 & ED &   5.798380 & W13 \\
n028235 & S742000286 & 12:21:35.82 & -88:00:14.50 &  12.20 &  12.17 &  11.71 &  12.26 & ES &   1.892910 & W13 \\
n029044 & S742000186 & 13:28:28.82 & -87:53:09.20 &   9.52 &  11.98 &  10.38 &  10.82 & MP &  11.564653 & W13 \\
n030008 & S3Y9000236 & 10:25:53.99 & -87:53:40.80 &   9.83 &   \nd  &  10.63 &  10.96 & MP &  20.978237 & W13 \\
n036260 & S742000480 & 12:46:26.19 & -88:14:42.11 &  11.61 &  11.49 &  10.63 &  11.51 & PR & 102.304580 & New \\
n038457 & S742000399 & 14:11:08.38 & -88:03:32.83 &   9.77 &  11.24 &  11.55 &  11.16 & PR &  15.832869 & New \\
n039664 & S742000504 & 13:23:49.26 & -88:16:04.30 &  12.43 &  12.78 &  11.92 &  12.58 & ES &   2.510726 & W13 \\
n040035 & S3Y9000616 & 11:52:51.12 & -88:23:28.90 &   9.21 &  11.17 &   9.80 &   9.98 & MP &  17.086311 & W13 \\
n040066 & S3Y9000570 & 11:08:22.53 & -88:21:01.33 &   8.64 &   9.94 &   8.18 &   9.13 & IR &   \nd      & New \\
n040892 & S742000553 & 13:12:16.91 & -88:20:34.58 &  11.97 &  11.97 &  11.46 &  11.73 & IR &   \nd      & New \\
n041768 & S3Y9000003 & 09:24:49.40 & -88:00:09.50 &  11.14 &  11.89 &  10.05 &  11.30 & IR &   \nd      & New \\
n042221 & S3Y9000527 & 10:01:21.80 & -88:13:30.80 &  11.88 &  11.51 &  10.67 &  11.55 & EC &   0.652314 & W13 \\
n043309 & S742026828 & 13:34:47.76 & -88:21:15.76 &  11.58 &  12.40 &  11.28 &  11.74 & IR &   \nd      & New \\
n043406 & S3YN000632 & 08:39:40.85 & -87:39:02.30 &  12.21 &   \nd  &  12.58 &  12.67 & ED &   7.165431 & W13 \\
n045983 & S742000656 & 12:12:56.28 & -88:34:11.60 &  10.77 &  11.70 &  10.83 &  11.08 & MP &   9.972472 & W13 \\
n047552 & S3YB000429 & 11:17:00.40 & -88:35:36.40 &   9.35 &  12.08 &  10.21 &  10.87 & LT &   \nd      & W13 \\
n052325 & S74E000072 & 15:44:44.17 & -87:46:35.40 &   8.32 &   9.09 &   \nd  &   8.76 & IR &   \nd      & W13 \\
n052442 & S743000136 & 15:01:18.06 & -88:13:44.00 &   9.07 &  12.30 &  11.86 &  11.19 & IR &   \nd      & New \\
n052690 & S743000171 & 14:50:47.94 & -88:19:40.37 &   8.84 &  11.82 &  10.99 &  10.60 & IR &   \nd      & New \\
n053736 & S3YB000143 & 08:57:31.35 & -88:18:29.95 &   9.88 &  10.59 &  10.04 &  10.93 & IR &   \nd      & New \\
n054284 & S3YB000225 & 09:29:39.14 & -88:30:06.40 &  11.78 &  12.16 &  11.58 &  12.00 & RL &   0.621863 & W13 \\
n055150 & S3YB000458 & 10:04:40.34 & -88:40:25.50 &   9.06 &   9.25 &   8.95 &   9.23 & PR &   0.092508 & W13 \\
n055656 & S741000043 & 15:32:36.94 & -88:07:43.28 &   9.82 &  11.01 &  10.10 &  10.74 & IR &   \nd      & New \\
n057314 & S3YB000128 & 08:39:19.42 & -88:13:56.39 &   9.63 &  13.14 &  12.56 &  11.90 & IR &   \nd      & New \\
n057617 & S743000094 & 13:50:03.38 & -88:46:13.20 &  10.08 &  11.08 &  10.42 &  10.50 & MP &  15.167192 & W13 \\
n057725 & S3YB000482 & 10:01:18.94 & -88:44:36.80 &  10.08 &  11.11 &  10.24 &  10.41 & MP &  43.205799 & W13 \\
n058002 & S743000311 & 14:29:04.38 & -88:38:43.70 &  12.07 &  12.54 &  11.88 &  12.59 & RL &   0.646577 & W13 \\
n058442 & S3YB000199 & 08:53:45.85 & -88:26:33.00 &  12.41 &  13.14 &  12.27 &  12.83 & PR &   0.258295 & W13 \\
n059543 & S3YB000243 & 09:03:59.29 & -88:33:07.60 &  11.40 &  11.80 &  11.28 &  11.66 & ES &   0.873857 & W13 \\
n060041 & S743000153 & 15:35:01.14 & -88:16:11.90 &  12.99 &  14.76 &  13.29 &  14.36 & LT &   \nd      & W13 \\
n060076 & S743000364 & 14:14:12.69 & -88:45:41.26 &   8.98 &   9.42 &   9.24 &   9.31 & IR &   \nd      & New \\
n060789 & S741000025 & 15:59:17.54 & -88:00:42.50 &  14.19 &  14.49 &   \nd  &  14.52 & ED &   6.853790 & W13 \\
n062144 & S743000115 & 13:53:18.49 & -88:54:14.60 &  12.82 &  12.96 &  12.23 &  12.83 & EC &   0.266903 & W13 \\
n062640 & S3YB000253 & 08:46:12.64 & -88:33:42.90 &  11.97 &  13.07 &  12.14 &  12.39 & EC &   0.267127 & W13 \\
n063743 & S740000060 & 12:36:24.67 & -89:04:02.70 &  10.97 &  12.65 &  11.34 &  11.64 & MP &  25.161758 & W13 \\
n068660 & S3YB009826 & 08:40:28.89 & -88:47:00.40 &  13.72 &  13.87 &  13.28 &  14.35 & ES &  13.024607 & W13 \\
n077969 & S3Y8000195 & 08:12:34.42 & -89:02:15.07 &  12.46 &  11.97 &  12.08 &  12.38 & IR &   \nd      & New \\
n078169 & S3Y8000251 & 09:35:54.44 & -89:19:28.38 &  10.04 &  11.65 &  11.44 &  10.22 & IR &   \nd      & New \\
n080649 & S3YA000502 & 07:01:38.41 & -88:17:02.90 &   9.60 &  11.79 &  10.33 &  10.55 & MP &  23.466223 & W13 \\
n082370 & S3YB000086 & 07:23:35.24 & -88:51:06.97 &   9.03 &   8.55 &   8.52 &   8.95 & PR &   1.618226 & New \\
n083359 & S3Y8000078 & 07:43:54.49 & -89:07:37.30 &  12.50 &  13.08 &  12.08 &  12.68 & EC &   0.797910 & W13 \\
n084427 & S3Y8000109 & 07:54:37.65 & -89:15:40.90 &   9.75 &  12.18 &  10.71 &  10.97 & IR &   \nd      & W13 \\
n085005 & S740000411 & 16:02:18.53 & -89:19:14.30 &  12.73 &  12.70 &  11.79 &  12.54 & IR &   \nd      & New \\
n086263 & S3YA000492 & 06:40:47.15 & -88:15:21.30 &  11.70 &  12.08 &  11.62 &  11.93 & EC &   0.438659 & W13 \\
n087149 & S743000500 & 17:08:55.01 & -88:44:32.29 &  13.66 &  14.11 &  13.63 &  14.33 & PR &  28.912195 & New \\
n088653 & S3YA000336 & 06:28:42.76 & -88:02:41.70 &  12.34 &  12.80 &  12.03 &  12.54 & ED &   7.254001 & W13 \\
n090586 & S740000342 & 17:15:45.51 & -89:00:42.80 &  10.78 &  10.85 &  10.66 &  10.87 & PR &   0.022641 & W13 \\
n090919 & S741000489 & 17:36:45.98 & -88:14:10.50 &  11.31 &  11.38 &  11.08 &  11.36 & PR &   0.076166 & W13 \\
n095083 & S741000460 & 17:51:13.16 & -88:09:48.80 &  10.65 &  13.35 &  11.41 &  11.91 & MP &  30.763140 & W13 \\
n096554 & S740000469 & 17:05:16.14 & -89:51:43.80 &   9.68 &  11.90 &  10.46 &  10.72 & MP &  26.623756 & W13 \\
n097333 & S741000378 & 17:59:00.73 & -88:01:32.90 &  11.76 &  14.28 &  12.78 &  13.36 & MP &  38.853951 & W13 \\
n099159 & S0SG000328 & 05:52:16.66 & -89:00:35.42 &   9.19 &  11.35 &   9.98 &  10.35 & PR &  20.460938 & New \\
n099251 & SA9S000144 & 18:22:33.29 & -89:36:22.90 &  11.39 &  12.34 &  11.52 &  12.00 & PR &   2.853950 & W13 \\
n100083 & SA9U000383 & 18:08:15.09 & -88:18:02.90 &  10.83 &  11.32 &  10.87 &  11.14 & MP &   2.842765 & W13 \\
n102641 & S0SH000215 & 05:47:08.05 & -87:51:00.20 &  10.23 &  10.07 &  10.15 &  10.49 & GD &   0.606546 & W13 \\
n104524 & SA9V000050 & 18:30:57.87 & -88:43:17.50 &   9.86 &  10.20 &   9.80 &   9.98 & ED &   9.925551 & W13 \\
n104943 & SA9U000438 & 18:29:03.93 & -88:32:31.90 &  13.27 &  13.67 &  12.63 &  13.22 & RL &   0.573044 & W13 \\
n105244 & S0SG000150 & 00:20:19.58 & -89:48:38.00 &   9.43 &  11.63 &  10.22 &  10.31 & MP &  10.923423 & W13 \\
n106372 & SA9U000442 & 18:35:32.31 & -88:33:47.92 &  11.97 &  12.54 &  11.98 &  12.27 & PR &   0.573259 & W11 \\
n107579 & SA9V000058 & 18:41:51.83 & -88:46:11.42 &   9.45 &  13.03 &  12.84 &  11.67 & IR &   \nd      & New \\
n110665 & S0SH000448 & 05:15:49.62 & -88:17:51.50 &  10.21 &  12.23 &  10.76 &  10.97 & MP &  11.462763 & W13 \\
n110942 & SA9S000107 & 19:55:16.09 & -89:18:10.62 &  10.28 &  11.90 &  11.79 &  11.61 & IR &   \nd      & New \\
n112694 & SA9V000073 & 19:17:53.08 & -88:51:11.20 &  11.81 &  12.60 &  11.60 &  12.05 & EC &   0.372068 & W13 \\
n113486 & SA9S000068 & 19:48:57.10 & -89:07:14.30 &  10.86 &  10.94 &  10.62 &  11.04 & PR &   4.842585 & W13 \\
n115348 & SA9U000331 & 18:58:35.51 & -88:12:55.20 &  11.71 &  13.12 &  11.80 &  12.19 & MP &  62.185490 & W13 \\
n118528 & SA9S000384 & 23:35:11.99 & -89:27:51.80 &  14.48 &  12.59 &  11.88 &  12.26 & RL &   0.465790 & W13 \\
n122836 & SAA5000323 & 19:02:33.97 & -87:35:30.20 &   8.86 &  10.18 &   9.32 &   9.32 & IR &   \nd      & W13 \\
n123187 & SA9S000168 & 20:57:31.47 & -89:03:50.30 &  12.34 &  11.17 &  10.22 &  10.39 & ED &   1.857642 & W13 \\
n123522 & SA9U000336 & 19:27:57.26 & -88:13:26.20 &   8.81 &  13.01 &   9.91 &  11.19 & IR &   \nd      & W13\ \\
n123706 & S0SG000092 & 01:23;01.27 & -89:17:09.40 &  13.60 &  14.22 &  12.99 &  13.57 & DS &   0.193489 & W13 \\
n123782 & S0SH000485 & 04:20:11.85 & -88:25:03.50 &  12.62 &  13.02 &  12.28 &  13.11 & DS &   0.197741 & W13 \\
n124517 & S0SG000093 & 00:52:40.76 & -89:17:32.40 &  13.84 &  12.49 &  12.28 &  12.84 & EC &   0.292944 & W13 \\
n131494 & SA9S000300 & 22:17:44.44 & -89:01:38.10 &   9.45 &  12.83 &  10.73 &  11.41 & IR &   \nd      & W13 \\
n137559 & SAA5000417 & 19:50:26.13 & -87:44:50.70 &  12.84 &  13.14 &  12.63 &  12.68 & EC &   0.416436 & W13 \\
n138555 & S0SG000018 & 00:31:15.83 & -88:55:17.90 &  10.79 &  11.42 &  10.95 &  11.06 & MP &   9.422044 & W13 \\
n141342 & SA9V000172 & 21:10:19.88 & -88:27:33.98 &   8.60 &   9.63 &   8.65 &   8.53 & IR &   \nd      & New \\
n142981 & S0SJ000161 & 02:41:54.21 & -88:26:02.90 &  10.98 &  11.75 &  11.09 &  11.34 & MP &  11.204593 & W13 \\
n143160 & SA9U000449 & 20:35:13.18 & -88:07:55.70 &   8.98 &  11.24 &  11.36 &  10.76 & IR &   \nd      & New \\
n143876 & SA9U000450 & 20:41:00.36 & -88:08:13.92 &  12.27 &  12.70 &  12.16 &  12.17 & IR &   \nd      & New \\
n145960 & S0SJ000031 & 01:51:34.69 & -88:33:26.90 &  12.01 &  12.81 &  12.11 &  12.44 & MP &  10.882430 & W13 \\
n148233 & SAA5000503 & 20:28:30.07 & -87:46:16.50 &  11.81 &  12.52 &  11.82 &  12.05 & ED &   2.192580 & W13 \\
n148910 & S0SJ000041 & 00:37:50.59 & -88:37:31.19 &   9.15 &  11.25 &  11.28 &  11.19 & IR &   \nd      & New \\
n149414 & S0SJ000002 & 03:00:33.53 & -88:02:59.20 &   9.98 &  13.38 &  11.39 &  12.14 & LT &   \nd      & W13 \\
n152261 & S0SI000373 & 01:53:25.61 & -88:20:58.78 &   9.39 &  11.25 &  11.76 &  11.45 & IR &   \nd      & New \\
n152437 & S0SI000269 & 02:42:27.87 & -88:04:22.50 &   9.45 &  12.25 &  10.53 &  11.13 & MP &  44.303307 & W13 \\
n153006 & S0SI000438 & 02:12:56.05 & -88:13:52.50 &  10.03 &  11.94 &  10.69 &  10.75 & MP &  12.116866 & W13 \\
n155317 & S0SI000338 & 01:55:46.26 & -88:14:41.57 &  12.07 &  12.59 &  12.27 &  12.27 & IR &   \nd      & New \\
n155320 & S0SI000374 & 00:47:25.67 & -88:23:31.13 &   8.82 &  10.19 &   9.62 &   9.65 & IR &   \nd      & New \\
n157069 & S0SI000391 & 00:15:15.02 & -88:23:35.48 &  12.95 &  12.62 &  12.39 &  12.85 & IR &   \nd      & New \\
n159243 & S0SI000372 & 00:00:50.89 & -88:19:41.90 &   9.44 &   9.65 &   9.45 &   9.65 & PR &   0.114436 & W13 \\
n162294 & S0SI000329 & 00:08:43.43 & -88:13:48.40 &  10.07 &  12.44 &  10.91 &  11.20 & IR &   \nd      & W13 \\
n164527 & SA9T000310 & 23:15:35.46 & -88:07:33.80 &  10.78 &  12.58 &  11.26 &  11.53 & MP &  24.945468 & W13 \\
n165516 & S0SI000292 & 00:23:44.62 & -88:08:02.44 &   8.92 &  11.25 &  10.27 &  10.25 & IR &   \nd      & New \\
n168446 & SAA6000034 & 21:47:16.29 & -87:39:06.60 &  12.76 &  13.11 &   \nd  &   \nd  & RL &   0.458059 & W13 \\
n171256 & SA9T000182 & 23:16:45.97 & -87:54:16.60 &  11.95 &  13.04 &  12.21 &  12.43 & MP &   8.754028 & W13 \\
n177534 & S0SI000101 & 00:01:16.84 & -87:44:02.90 &  11.99 &  11.99 &  11.41 &  12.04 & ED &   9.458450 & W13 \\
n863059 & S3Y8000125 & 06:49:54.20 & -89:21:58.80 &   9.78 &  11.45 &  10.05 &  10.35 & IR &   \nd      & W13 \\
n897790 & SA9V000415 & 22:23:40.80 & -88:53:42.90 &   9.78 &  10.01 &   9.78 &   9.94 & GD &   0.521773 & W13 \\
p09-001 & S0SJ000117 & 03:29:35.78 & -88:14:49.52 &   \nd  &   8.68 &   8.20 &   8.39 & IR &   \nd      & New \\
p09-002 & S0SH000437 & 04:49:04.42 & -88:16:16.72 &   \nd  &   8.00 &   6.43 &   6.74 & PR &  22.166017 & AAVSO \\
p09-003 & S3Y9000510 & 11:20:32.71 & -88:14:59.89 &   \nd  &   9.26 &   7.59 &   8.09 & PR &  16.624512 & AAVSO \\
p09-004 & S740000291 & 14:35:26.74 & -89:46:18.19 &   \nd  &   7.58 &   5.92 &   6.32 & PR &  60.452774 & AAVSO \\
p09-005 & S743000188 & 14:59:52.13 & -88:22:35.26 &   \nd  &   9.34 &   8.55 &   8.63 & ES &   6.072881 & New \\
p09-006 & SA9V000311 & 21:14:24.35 & -88:56:02.00 &   \nd  &   8.20 &   8.73 &   8.40 & IR &   \nd      & New \\
p09-007 & SA9V000407 & 22:45:37.21 & -88:49:06.17 &   \nd  &   6.79 &   6.78 &   6.80 & DS &   0.122194 & AAVSO
\enddata
\tablenotetext{1}{W11 - \citealt{Wang2011}; W13 - \citealt{Wang2013}; AAVSO - American Association of Variable Star Observers; New - no current catalog entry}
\end{deluxetable*}

\end{document}